\documentclass[mypaper,7pt,twoside]{CoAst}
\usepackage{epsf,graphicx,fancyhdr}
\renewcommand{\chaptermark}[1]{\markboth{#1}{}}

\newcommand{\apj}{ApJ}
\newcommand{\mnras}{MNRAS}

\newcommand{\kopf}{\small\itshape Comm. in Asteroseismology \\ Contribution to the Proceedings of the Wroclaw HELAS Workshop, 2008}

\newcommand{\Authors}[1]{\begin{center}\normalsize\bf\sf #1 \end{center}}

\renewcommand{\author}[1]{\begin{center}\normalsize\bf\sf #1 \end{center}}
\newcommand{\Address}[1]{\begin{center}\small\sf #1 \end{center}}

\newcommand{\Session}[1]{{\vspace{3mm}\small \noindent  \hspace*{3mm} Session: } #1 \normalsize}

\newcommand{\Objects}[1]{{\vspace{0mm}\small \noindent  \hspace*{3mm} Individual Objects: } \small #1 \normalsize}

	\newcommand{\twoA}{\small MODES - extracting eigenmode frequencies \newline}

\renewenvironment{abstract}{\section*{Abstract}\normalsize\sf}{}
\newcommand{\References}[1]{\begin{flushleft}{\large References\\}\vspace*{2mm}\small #1 \end{flushleft}}

\newcommand{\chapterCoAst}[2]{\chapter[\sf\normalsize #1\\ \footnotesize \hspace*{5mm}by #2 \sf\normalsize][]{#1\\}\rhead[\fancyplain{}{\sf\footnotesize \center{#1}}]{\fancyplain{}{\sffamily\thepage}}\lhead[\fancyplain{\kopf}{\sffamily\thepage}]{\fancyplain{\kopf}{\sf\footnotesize \center{#2}}}}

\newcommand{\figureCoAst}[5]{\begin{figure}[#4]
\centering
\includegraphics*[#5]{#1}
\caption{#2}
\label{#3}
\end{figure}}

\renewcommand{\chaptername}{}

\newcommand{\acknowledgments}[1]{\vspace*{5mm}\noindent  \textbf{Acknowledgments.} #1}

\def\rfr{\smallskip\par\noindent
        \hangindent=7truemm
        \hangafter=1}

\begin{document}
\sf

\chapterCoAst{Application of the Trend Filtering Algorithm in 
              the search for multiperiodic signals}
{G.\,Kov\'acs and G.~\'A.\,Bakos} 
\Authors{G.\,Kov\'acs$^{1}$, and G.~\'A.\,Bakos$^{2}$} 
\Address{
$^1$ Konkoly Observatory, P.O. Box 67, Budapest H-1125, Hungary\\
$^2$ Harvard-Smithsonian Center for Astrophysics, 60 Garden Street, Cambridge, MA 02138, USA\\
}

\noindent
\begin{abstract}
During the past few years the Trend Filtering Algorithm (TFA) 
has become an important utility in filtering out 
time-dependent systematic effects in photometric databases for 
extrasolar planetary transit search. Here we present the extension 
of the method to multiperiodic signals and show the high efficiency 
of the signal detection over the direct frequency analysis on 
the original database derived by today's standard methods (e.g., 
aperture photometry). We also consider the (iterative) signal 
reconstruction that involves the proper extraction of the systematics. 
The method is demonstrated on the database of fields observed by 
the HATNet project. A preliminary variability statistics suggests
incidence rates between $4$ and $10$\% with many (sub)mmag amplitude 
variables. 
\end{abstract}

\Session{\twoA}
\Objects{} 

%
%
\section*{Introduction}
The Trend Filtering Algorithm (TFA) has been routinely used during the 
past several years in the search for transiting extrasolar planets 
within the HATNet\footnote{{\bf H}ungarian-made {\bf A}utomated 
{\bf T}elescope {\bf Net}work \hfill \break 
{\em http://cfa-www.harvard.edu/~gbakos/HAT/}} project (Bakos et.~al 2004). 
The goal of this post-processing method is to filter out systematics/trends 
from the photometric time series. The presence of these effects is due 
to sub-optimal observing conditions, data acquisition and reduction; e.g., remaining differential extinction, 
distorted, position- and time-dependent point spread function, 
astrometric errors, etc. Although wide field observations are the 
ones most affected by systematics, the fingerprints of these 
perturbations are always present in nearly all photometric observations 
(in surveys, such as MACHO -- Alcock et al.~2000, or in individual 
object followup observations by small field-of-view telescopes 
-- Kov\'acs \& Bakos 2007). 

Effects of systematics have not been considered in the past too closely, 
since, relatively speaking, they play less important role in large 
amplitude variables, and most of the earlier investigations focused on 
specific classes of stars without paying attention to the ``constant'' 
stars, displaying the systematics in the most obvious way (due to the 
lack of more prominent physical variations). This situation has changed 
with the advent of the microlensing surveys, when it has become clear 
that more sophisticated image processing tools, such as the image 
subtraction method (ISIS, see Alard \& Lupton 1998) are needed to 
disentangle weak signals and systematics when searching for variables 
in crowded fields. While the above differential image analysis works 
on the images (snapshots of the full photometric time series), TFA 
(Kov\'acs, Bakos \& Noyes 2005; hereafter KBN) and SysRem (Tamuz, Mazeh 
\& Zucker 2005) attempt to utilize the information available in the 
full time history of the light curves. 

In the following we briefly summarize the main steps of the algorithm, 
extend the method to multiperiodic time series, demonstrate the 
effectiveness of the method by various tests and perform a brief 
variability survey on $10$ HATNet fields.    
 
%
%
\section*{TFA with multiperiodic signal reconstruction}
Here we briefly summarize the main assumptions and formulae of TFA. 
The interested reader is referred to KBN and Kov\'acs \& Bakos (2007) 
for additional details. 

The basic assumptions are the following: (i) systematics are present 
in several/many objects in the field (i.e., TFA template selection is 
possible); (ii) trends in any target are linearly decomposable by using 
some subset (template) of time series available in the field; (iii) the 
observed time series is trend- and noise-dominated\footnote{
This property is used only in the frequency search. For signal 
reconstruction the full time series model is used, including the 
hidden signal component.}; (iv) there is a common time base for the 
large majority of objects. After selecting a set of templates 
(\{$X_{\rm k}(i), k=1,2,...,M; i=1,2,...,N$\} -- with $k$ being the 
template and $i$ is the time index), for each target we compute a filter 
$F(i)$ 
%
%
\begin{eqnarray}
F(i) = \sum_{\rm k=1}^{\rm M} c_{\rm k} X_{\rm k}(i)\hskip 2mm ,
\end{eqnarray}
where the coefficients \{$c_{\rm k}; k=1,2,...,M$\} are derived 
from the following condition for each observed time series 
\{$Y(i); i=1,2,...,N$\} 
%
%
\begin{eqnarray}
\sum_{\rm i=1}^{\rm N} [Y(i)-A(i)-F(i)]^2 = min\hskip 2mm .
\end{eqnarray}
Here the function \{$A(i); i=1,2,...,N$\} is either constant, 
or is the trend- and noise-free signal, to be found iteratively 
in the signal reconstruction phase. For single- and multiperiodic 
signals, when the Fourier representation of the signal is 
adequate, we can perform signal reconstruction without iteration. 
In this case the Fourier part is included in $F(i)$ 
%
%
\begin{eqnarray}
F(i) & = & \sum_{\rm k=1}^{\rm M} c_{\rm k} X_{\rm k}(i)  
       +   \sum_{\rm j=1}^{\rm 2L} a_{\rm j} S_{\rm j}(i) \hskip 2mm ,
\end{eqnarray}
where \{$S_{\rm j}(i); j=1,2,...,{\rm 2L}; i=1,2,...,N$\} are the 
Fourier components ({\em sine} and {\em cosine} functions) with 
${\rm L}$ different frequencies and \{$a_{\rm j}$\} phase-dependent 
amplitudes. The frequencies are determined from the analysis of a time
series derived by Eqs.~(1) and (2) with ``no signal'' assumption 
(i.e., with \{$A(i)=const$\}). Assuming that these frequencies 
approximate well the ones representing the noise- and trend-free time 
series, the advantage of Eq.~(3) is that it yields an {\em exact} 
solution in one step for signals of the form of 
{\em trend + Four.~comp.~+ noise}. If the signal has additional 
components (e.g., transients, transits) that are not well-represented 
by a finite Fourier sum, we should use a more complicated model and, 
as a consequence, an iterative scheme to obtain approximations for 
the signal components. We note that, in principle, iteration should 
be employed also if the non-sinusoidal components are absent, because 
the starting model from which we determine the frequencies is 
different from the one used in the reconstruction. However, based 
on our experience from the application of the ``no signal'' assumption 
in periodic transit search, the frequencies derived in this way are 
accurate enough, and there is no need for the very time-consuming 
iterative procedure in the frequency search.   

%
%
\section*{Tests, examples}
In KBN we presented several tests showing the signal detection 
capability of TFA on the early set of HATNet light curves, focusing 
mostly on the detection of periodic transits. Here we show some 
selected examples on the detection of sinusoidal (i.e., Fourier) 
signals on the latest, more extensive datasets. 

%
%
\figureCoAst{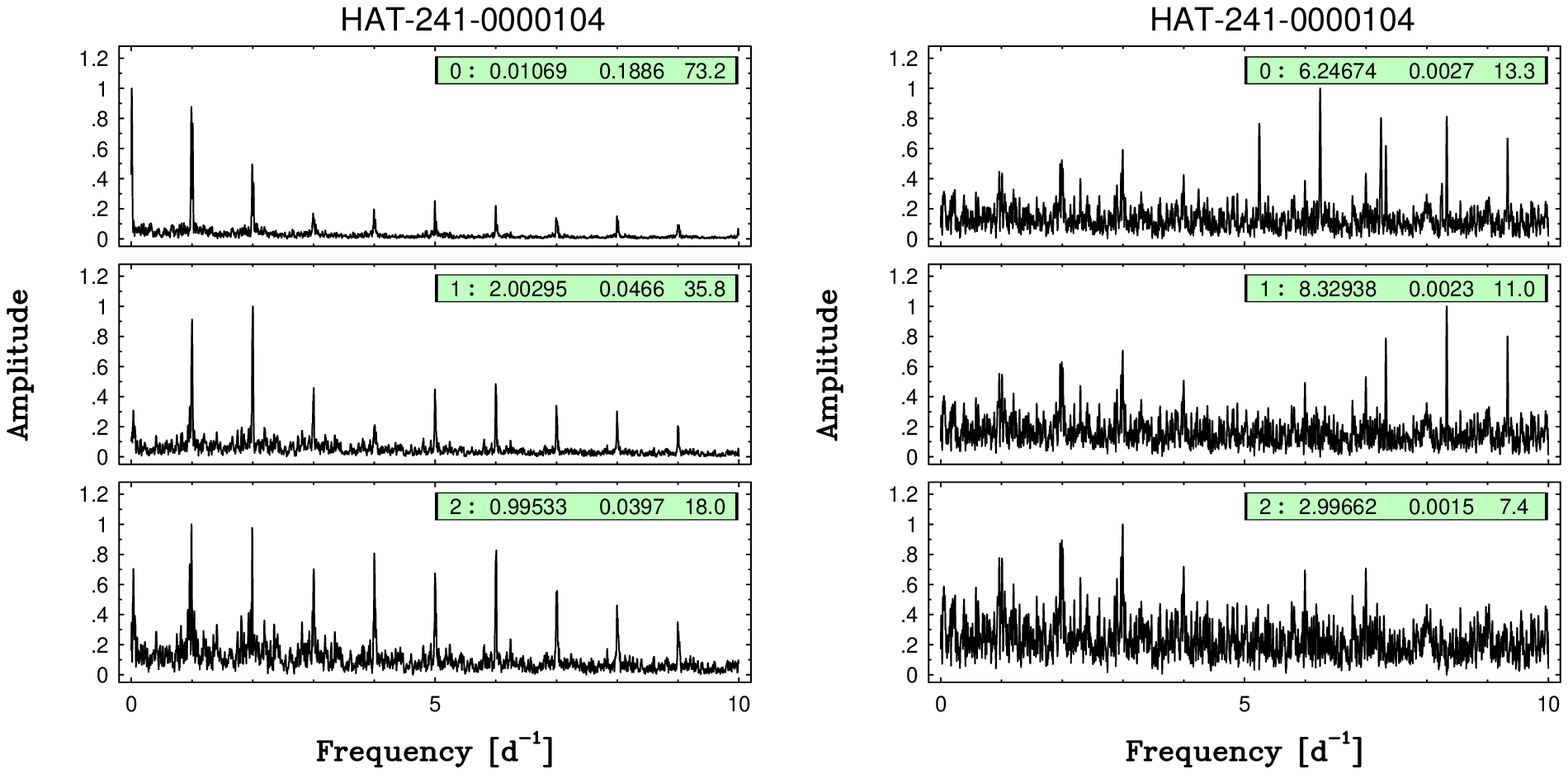}
{Panels on the left show the successive prewhitening of the 
{\em raw test time series} obtained by the injection of two 
sinusoidal components at 6.25 and 8.33~d$^{-1}$. Amplitudes 
are normalized, labels show the prewhitening cycle number, 
peak frequency, amplitude [mag] and signal-to-noise ratio. 
Simple Fourier prewhitening cannot recover the signal. 
Panels on the right show the result obtained by TFA filtering 
with $900$ templates. Both injected signal components are 
recovered with high significance.}
{fig1}{h}{clip,angle=0,width=105mm}

One of the questions that can be asked is why direct Fourier 
filtering is not used to clean up the data from systematics. 
The reason is threefold: (i) there are systematics (e.g., transients) 
for which Fourier representation is a rather bad one; (ii) we do not 
know {\em a priory} which component can be treated as trend and which 
one as signal; (iii) for the most common periodic (daily) systematics 
Fourier filtering is less stable, because of the gaps in the data 
with the same periodicity. Figure~1 demonstrates the inadequacy of 
the simple Fourier filtering. The injected low-amplitude signal 
remains completely hidden if we employ direct Fourier filtering. 
Although TFA filtering also leaves some trend in the data (see the 
peak in the bottom right panel at $3.0$~d$^{-1}$), its amplitude is 
$26$-times smaller than that of the highest peak in the direct Fourier 
filtering at the same stage of prewhitening.   

%
%
\figureCoAst{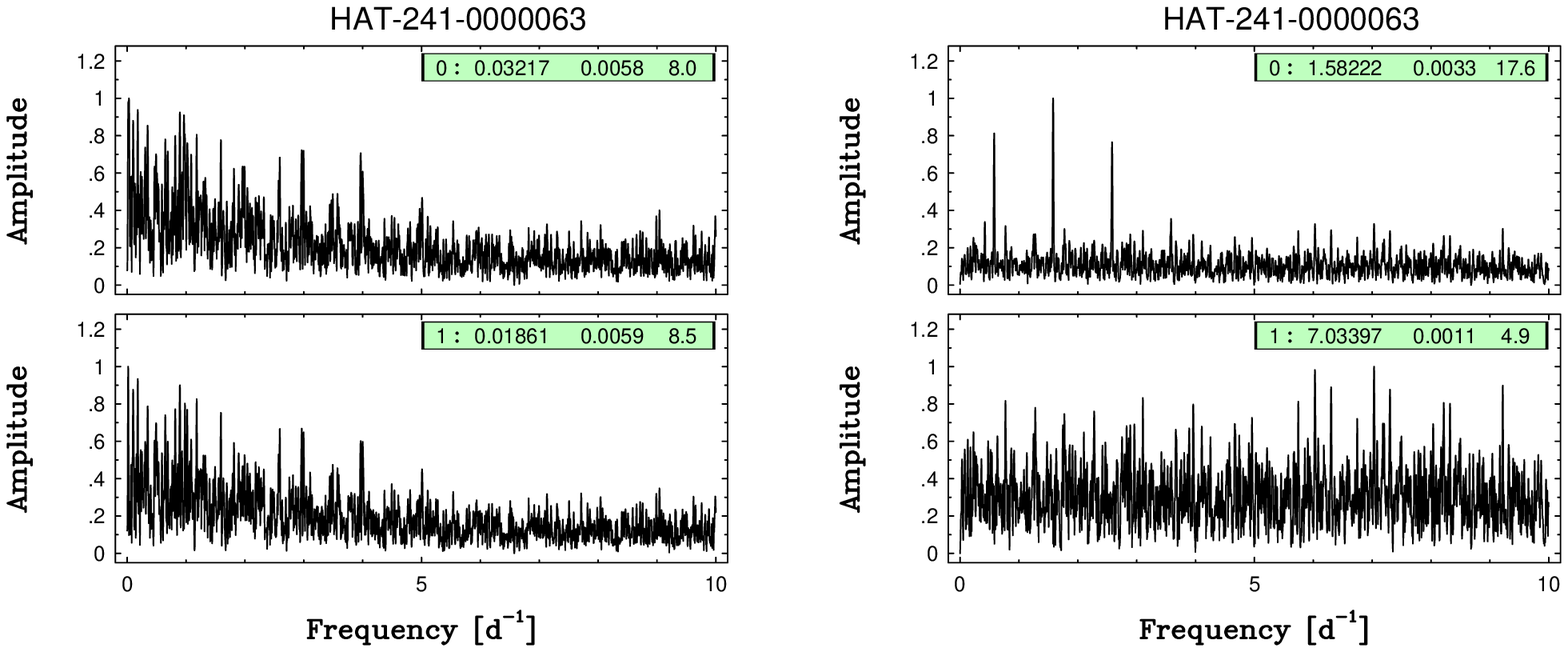}
{Example on a variable that is hidden in the raw time series 
(panels on the left) but becomes highly visible in the TFAd 
time series (panels on the right). Notation is as in Fig.~1.}
{fig2}{h}{clip,angle=0,width=105mm}
%

%
%
\figureCoAst{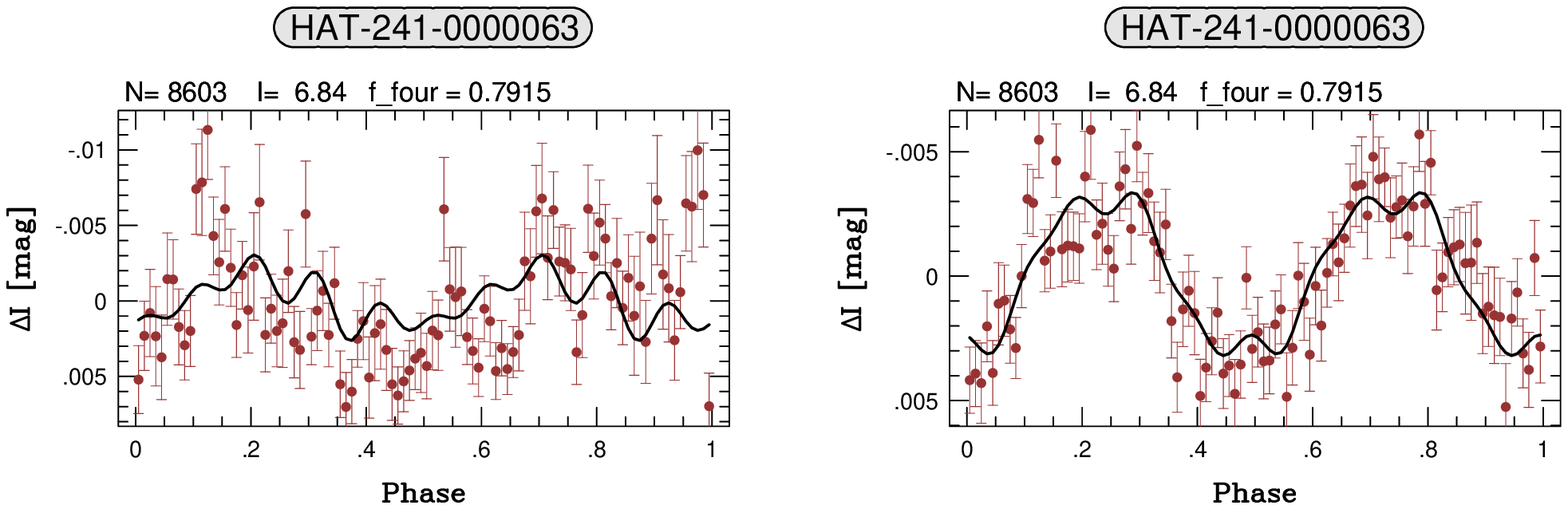}
{Folded/binned light curves with twice of the period of the variable 
shown in Fig.~2. Left: raw data, right: TFAd data. Headers from 
left to right: number of data points, average ``I'' magnitude, 
folding frequency in d$^{-1}$.}
{fig3}{h}{clip,angle=0,width=105mm}
%

%
%
\figureCoAst{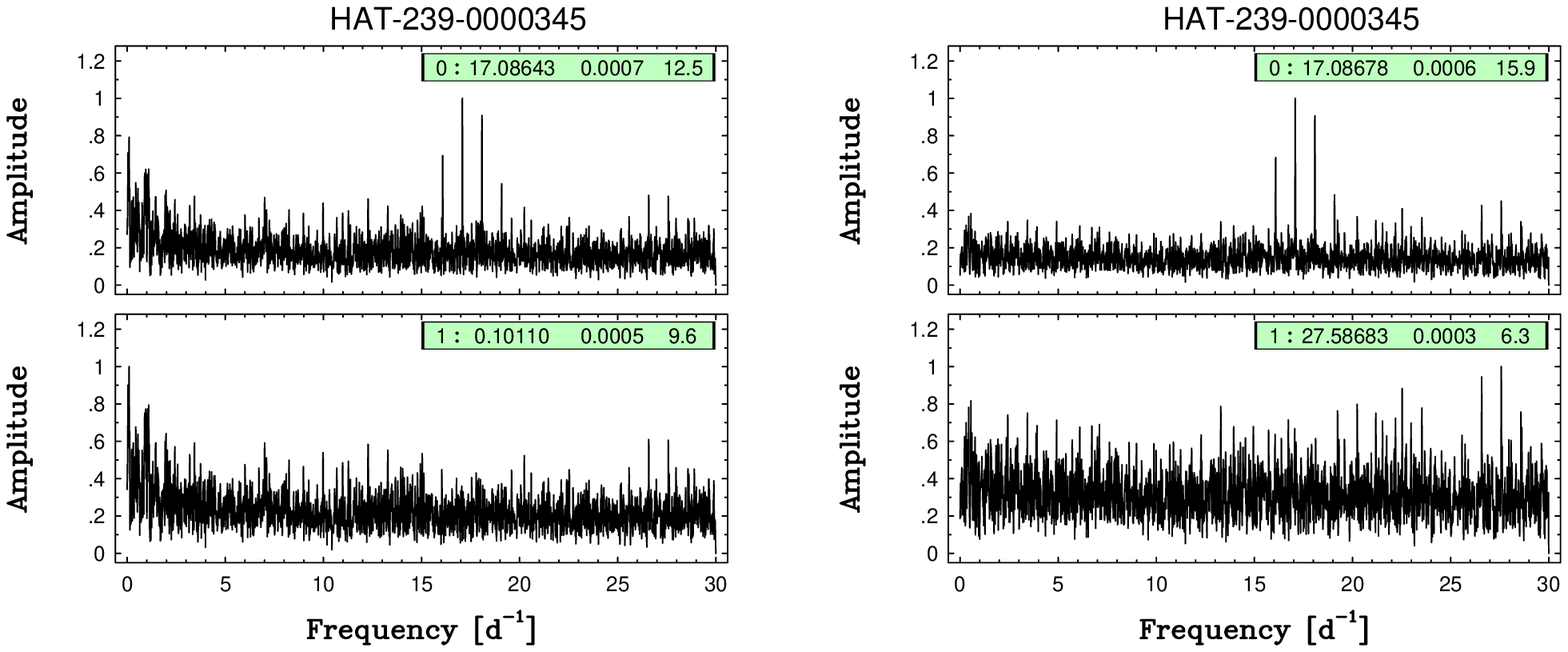}
{Example on a sub-millimag variable. The signal is detectable also 
in the raw time series (left) but is cleaner in the TFA filtered 
one (right). Notation is as in Fig.~2.}
{fig4}{h}{clip,angle=0,width=105mm}

Next, in Fig.~2 we show the frequency spectra of a real variable 
that has escaped detection in the original time series. The star 
is rather bright and therefore it is strongly affected by various 
saturation-related effects. These effects are also common in 
other bright stars in the field, so it is possible to filter 
them out by employing TFA. In Fig.~3 we also show the folded light 
curves to give another look at the difference between the raw and the 
TFA-reconstructed results. Finally, as an example of the detection 
capability on the HATNet database, in Fig.~4 we show the frequency 
spectra of a sub-millimag variable.

%
%
\section*{Brief HATNet variability statistics}
By using TFA post-processing, we have Fourier analyzed 10 HATNet 
fields in the $[0.0,20.0]$~d$^{-1}$ range and searched for variables 
with high significance (SNR$>10$) in the frequency spectra. The 
number of stars analyzed per field varies between 10000 and 25000, 
with 5000 to 11000 data points per object. The time spans covered 
by the observations are between 100 and 1000 days. The incidence 
rates of the variables are between 4 and 10\%. The number of 
sub-mmag variables changes from field-to-field, but it is typically 
in the order of 100. All these statistics are, of course, strong 
functions of the data quality, time span of the observations and 
sample of objects. The total number of objects analyzed is 169000, 
covering a magnitude range of $7<V<13$. The number of variables is 
9900. Some 12\% of these are sub-mmag variables. For comparison, 
in an effort to produce a variable input catalog for the Kepler 
fields, Pigulski et al. (2008) analyzed 250000 objects from 
the ASAS database. They found a variability rate of 0.4\%. This low 
incidence rate is not surprising if we consider that the average 
number of data points in these ASAS variables is only 100.

\acknowledgments{We thank for the support of the Hungarian Scientific 
Research Fund (OTKA, grant No. K-60750).
Work of G.~\'A.~B was supported by NSF fellowship AST-0702843. 
Operation of HATNet have been funded by NASA grants NNG04GN74G and
NNX08AF23G.
}

\References{
\rfr Alard, C. \& Lupton, R. H.~1998, \apj, 503, 325 
\rfr Alcock, C., Allsman, R., Alves, D. R. et al.~2000, \apj, 542, 257
\rfr Bakos, G. \'A., Noyes, R. W., Kov\'acs, G. et al.~2004, PASP, 116, 266
\rfr Kov\'acs, G., Bakos, G. \'A., \& Noyes, R. W.~2005, \mnras, 356, 557 (KBN)  
\rfr Kov\'acs, G., Bakos, G. \'A.~2007, ASP Conf. Ser., 366, 133
\rfr Pigulski, A., Pojmanski, G., Pilecki, B., \& Szczygiel, D.~2008, arXiv:0808.2558v2 
\rfr Tamuz, O., Mazeh, T. \& Zucker, S.~2005, \mnras, 356, 1466
}

\end{document}